\documentclass{aastex62}

\usepackage{booktabs}

\newcommand{\hydro}{\ion{H}{2}~}

\newcommand{\WISE}{\emph{WISE}~}
\newcommand{\co}{$^{12}$CO~}
\newcommand{\coiso}{$^{13}$CO~}

\received{}
\revised{}
\accepted{}

\submitjournal{ApJ}

\shorttitle{The Molecular Clouds of CTB 102}
\shortauthors{Marshall et al.}

\begin{document}

\title{High-Resolution Observations of the Molecular Clouds Associated with the Huge \hydro Region CTB 102}

\correspondingauthor{Sung-ju Kang}
\email{sjkang@kasi.re.kr}

\author[0000-0002-6661-0245]{Brandon Marshall}
\affiliation{Iowa State University \\
Dept. of Physics \& Astronomy, 2323 Osborne Dr. \\
Ames, IA 50011-3160, USA}

\author[0000-0002-5004-7216]{Sung-ju Kang}
\affiliation{Korea Astronomy and Space Science Institute \\
776, Daedeokdae-ro, Yuseong-gu \\
Daejeon, 34055, Republic of Korea}

\author[0000-0003-1539-3321]{C. R. Kerton}
\affiliation{Iowa State University \\
Dept. of Physics \& Astronomy, 2323 Osborne Dr. \\
Ames, IA 50011-3160, USA}

\author{Youngsik Kim}
\affiliation{Korea Astronomy and Space Science Institute \\
776, Daedeokdae-ro, Yuseong-gu \\
Daejeon, 34055, Republic of Korea}
\affiliation{Daejeon Observatory \\
213-48, Gwahak-ro, Yuseong-gu, Daejeon, 34128, Republic of Korea} 

\author{Minho Choi}
\affiliation{Korea Astronomy and Space Science Institute \\
776, Daedeokdae-ro, Yuseong-gu \\
Daejeon, 34055, Republic of Korea} 

\author[0000-0002-5016-050X]{Miju Kang}
\affiliation{Korea Astronomy and Space Science Institute \\
776, Daedeokdae-ro, Yuseong-gu \\
Daejeon, 34055, Republic of Korea}

\begin{abstract}
We report the first high-resolution (sub-arcminute) large-scale mapping \co and \coiso observations of the molecular clouds associated with the giant outer Galaxy \hydro region CTB~102 (KR 1). These observations were made using a newly commissioned receiver system on the 13.7-m radio telescope at the Taeduk Radio Astronomy Observatory. Our observations show that the molecular clouds have a spatial extent of $60 \times 35$ pc and a total mass of $10^{4.8} - 10^{5.0}$ M$_\sun$. Infrared data from \WISE and \emph{2MASS} were used to identify and classify the YSO population associated with ongoing star formation activity within the molecular clouds. We directly detect 18 class I/class II YSOs and six transition disk objects. Moving away from the \hydro region, there is an age/class gradient consistent with sequential star formation. The infrared and molecular-line data were combined to estimate the star formation efficiency (SFE) of the entire cloud as well as the SFE for various sub-regions of the cloud. We find that the overall SFE is between $\sim5$ -- $10$\%, consistent with previous observations of giant molecular clouds.  One of the sub-regions, region 1a, is a clear outlier, with a SFE of 17 -- 35\% on a 5 pc spatial scale. This high SFE is more typical for much smaller (sub-pc scale) star-forming cores, and we think region 1a is likely an embedded massive protocluster.
\end{abstract}


\section{Introduction} \label{intro}

In this paper we present the first \co and \coiso high-resolution (sub-arcminute) large-scale mapping observations of the molecular clouds associated with the CTB~102 \hydro region. These observations provide basic data about the clouds, such as their sizes and masses, and are combined with archival infrared data to explore ongoing star formation activity associated with the region. 

CTB~102 is an enormous \hydro region/bubble, with an estimated size of 100 -- 130 pc, located in the outer Galaxy (J2000: $21^\mathrm{h}12^\mathrm{m}21^\mathrm{s}$, $+52\degr28'59''$) at a distance $d = 4.3$~kpc \citep{arvidsson09}. It was first identified in low-resolution radio continuum surveys at 960 and 1420 MHz by \citet{WB1960} and \citet{KR80} respectively, and was imaged with $\sim 1'$ resolution at 1420 MHz as part of the Canadian Galactic Plane Survey \citep[CGPS][]{taylor03}. \citet{arvidsson09} used H89$\alpha$ radio recombination line observations to show that the extensive radio continuum structure visible in these surveys was all part of a single object with the primary structure having $V_{0\mathrm{,LSR}} = -62.66 \pm 0.05$ km s$^{-1}$ and most of the surrounding filaments having $|V - V_{0}| \lesssim 6$ km s$^{-1}$. They also determined that the total Lyman continuum photon emission rate from the region was $N_\mathrm{Ly} \geq (4.5 \pm 1.8) \times 10^{49}$ s$^{-1}$, consistent with a single early-type O star or with a cluster of several late-type O stars. 

In spite of its size, CTB~102 has not been studied at optical wavelengths because of both its distance and the fact it is hidden behind an extensive local region of high extinction \citep{lynds62,fitz68,simonson76}. At near- and mid-infrared wavelengths our best view of the region comes from the \emph{Wide-field Infrared Survey Explorer} \citep[\emph{WISE};][]{Wright} all-sky survey. A portion of the region was also observed by \emph{Spitzer} during its GLIMPSE360 warm-mission survey \citep{whitney08}. Comparable resolution data for molecular gas emission, which is essential for understanding star-formation activity associated with the \hydro region, was not obtained until this study. 

In \autoref{obs} we detail the new \co and \coiso observations as well as the archival data that has been collected for our study. In \autoref{analysis} the physical properties of the newly mapped molecular clouds are described, and the young stellar object (YSO) content of the clouds is determined using infrared data. The two studies then are combined to investigate the star-formation efficiency of the region. Finally, we discuss our findings and present our conclusions in \autoref{summary} and \autoref{conclusions}. 

\section{Observations} \label{obs}

\subsection{\co and \coiso Observations}

The CTB~102 \hydro region was observed using the Second Quabbin Optical Imaging Array in Taeduck Radio Astronomy Observatory (SEQUOIA-TRAO) receiver system on the TRAO 13.7-m radio telescope. The data were obtained between November 2016 and March 2017, during the first observing season after the SEQUOIA receiver array was relocated from the Five College Radio Astronomy Observatory (FCRAO) and adjusted to the TRAO system. Our data represent not just the first look at the CTB~102 region at \co and \textsuperscript{13}CO, but also some of the first data obtained by the SEQUOIA-TRAO system.

We observed \co (115.271 GHz, $J=1-0$) and \coiso (110.201 GHz, $J=1-0$) lines simultaneously using an On-The-Fly (OTF) observation mode in a $1\fdg5 \times 1\fdg5$ region in order to map the distribution of the molecular gas. Since the TRAO is a two local oscillator (LO) system, the spectral window has 4096$\times$2 channels in each 62.5 MHz bandwidth. In total, we observed 36 maps, each a 30$^{\prime}$ $\times$ 30$^{\prime}$ grid, centered on $l=93\fdg01$, $b=2\fdg73$. The RMS is $\sim0.3$~K, with a velocity resolution of 0.2 km s$^{-1}$. This compares favorably with the velocity resolution and RMS of the Outer Galaxy Survey \citep[OGS;][]{Heyer98} and the Galactic Ring Survey \citep[GRS;][]{grs06} both obtained at FCRAO: 0.98 km~ s$^{-1}$ and 0.65 km~s$^{-1}$ respectively, each with RMS $\sim0.5$ K. 

The observed data was reduced using {\sc{Otftool}} and the GILDAS software package {\sc{Class}}. The pointing of the observation was calibrated using the SiO $\nu=1, \ J=2-1$ maser line in the Orion IRc2 \citep{Baudry95}. The pointing observation was performed every two hours, with an average pointing error of $\sim6\arcsec$. The antenna temperature was corrected automatically for the effects of atmospheric attenuation with a standard chopper-wheel method.

The new radome of TRAO was installed in January 2017 to shield the antenna and receiver from the weather and other exterior environmental factors. Since the new radome affects the beam efficiency ($\eta_{\mathrm{mb}}$), we applied two different beam efficiencies at 115 GHz, for data acquired before ($\eta_{\mathrm{mb}}$ = 0.51) and after ($\eta_{\mathrm{mb}}$ = 0.46) the new radome installation. The full width at half maximum (FWHM) of the beam at 115 GHz (\textsuperscript{12}CO) is $45\arcsec$ and at 110 GHz (\textsuperscript{13}CO) is $48\arcsec$.

\begin{figure}[!htbp]	
\centering
\includegraphics[width=\columnwidth]{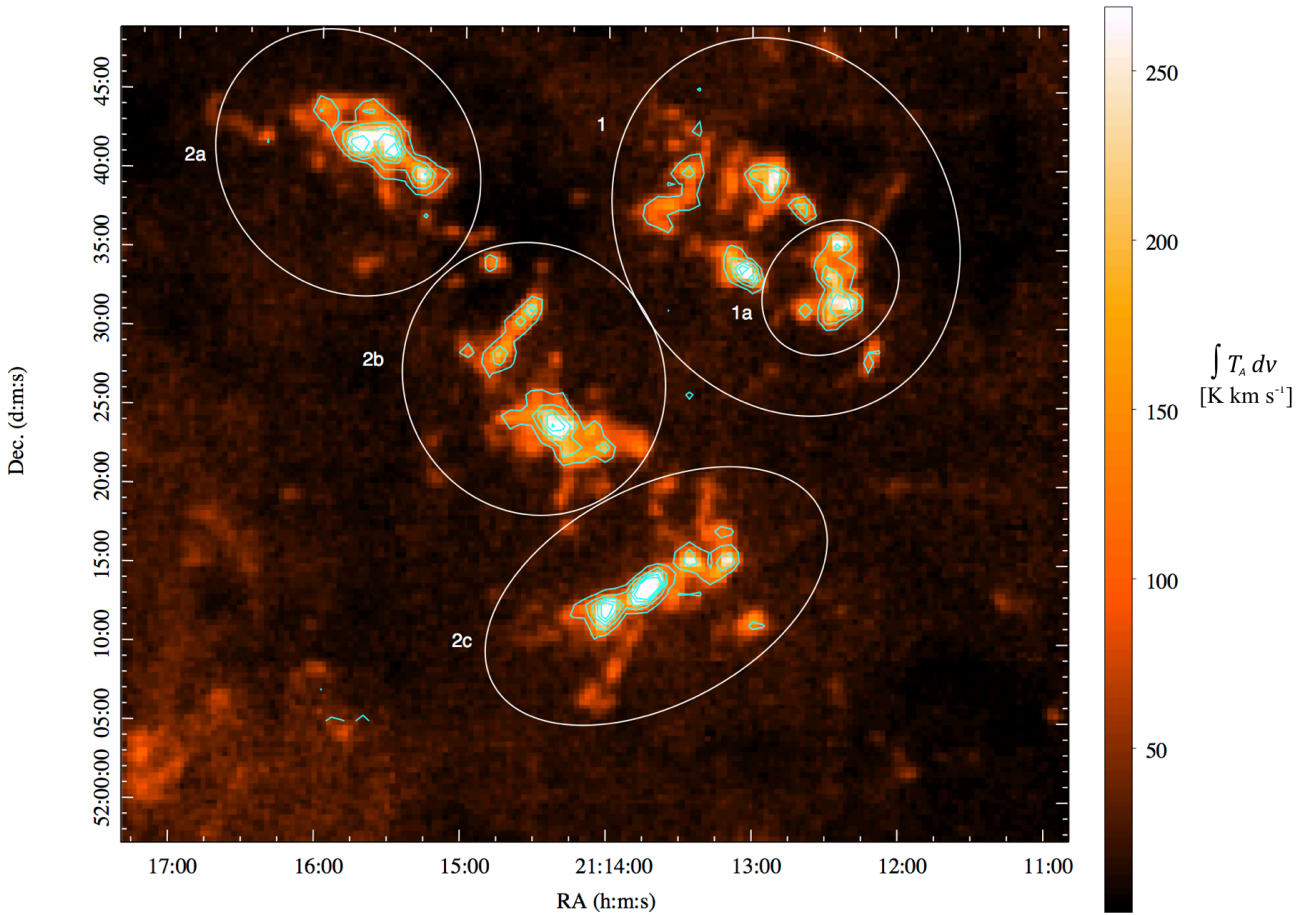}
    \caption{Integrated \co map of CTB~102. Integration was between $V_{\mathrm{LSR}} = -71$ km~s$^{-1}$ to $-53$ km~s$^{-1}$. White ellipses denote the four main subdivisions of the molecular cloud (1, 2a, 2b, and 2c) along with the smaller 1a region that has a high concentration of YSOs. Cyan contours show the integrated \coiso emission. Five contour levels were generated starting 3$\sigma$ above the median background. The morphology of the molecular cloud is very similar in both the \co and \coiso maps.}
    \label{fig:12CO}
\end{figure}

\begin{figure}[!htbp]
\centering
\includegraphics[width=\columnwidth]{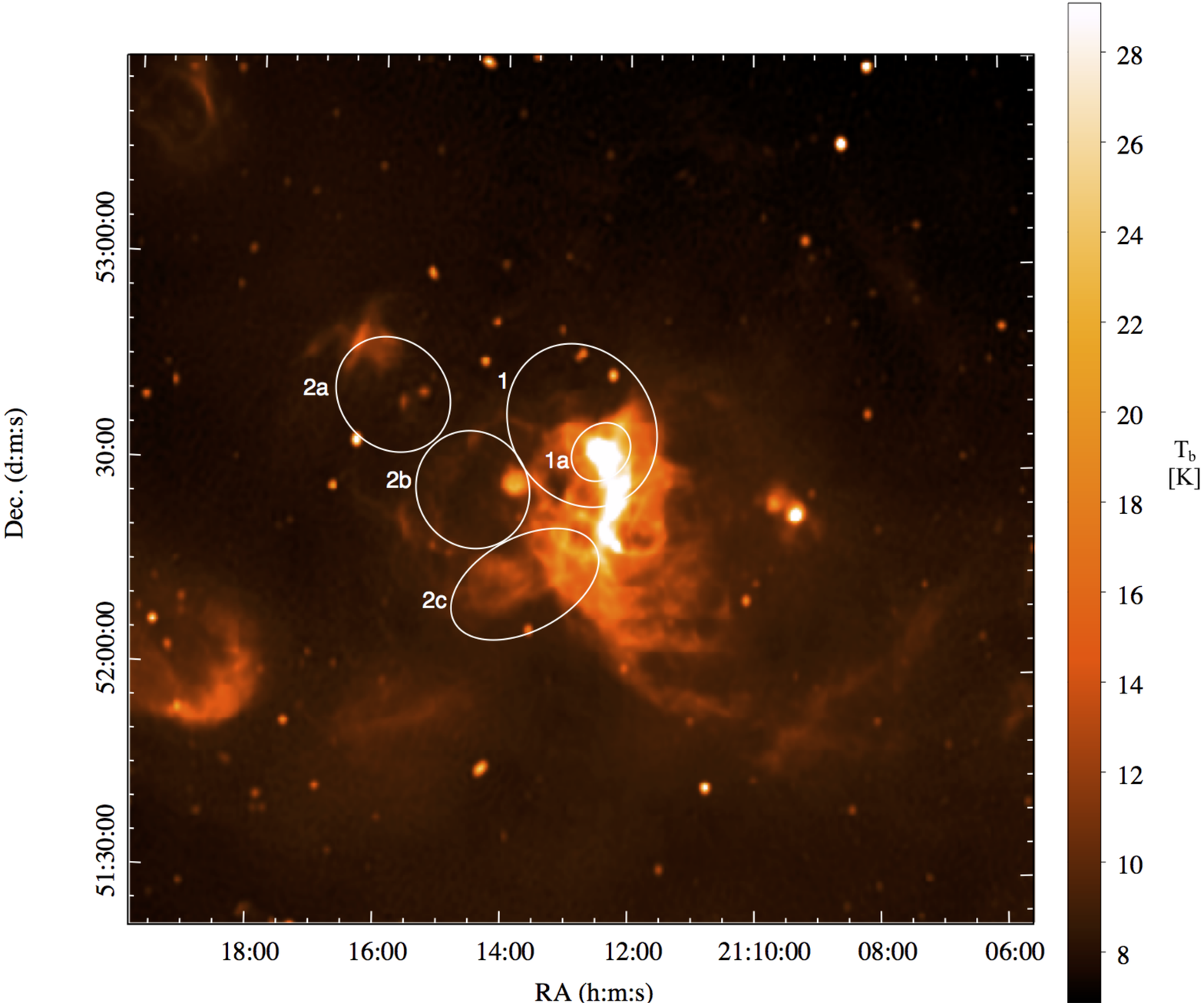}
	\caption{1420 MHz image of CTB 102 from the CGPS. The subdivisions of the molecular cloud from \autoref{fig:12CO} are shown. The color scale is brightness temperature (T\textsubscript{b}) in K.}
    \label{fig:1420}
\end{figure}

The integrated \co cube is shown in \autoref{fig:12CO}, with \coiso contours overlain. For reference, the location of the molecular clouds are also indicated on the CGPS 1420 MHz image of the region shown in \autoref{fig:1420}. 

\subsection{Archival Infrared Data}
\WISE scanned the entire sky in four bands centered at 3.4, 4.6, 12, and 22~$\mu$m with a 5$\sigma$ sensitivity of roughly 16.6, 15.6, 11.3, and 8.0 magnitudes respectively \citep{Wright}. We used the Infrared Science Archive (IRSA) to retrieved all of the sources in the \emph{WISE-}based AllWISE catalog within a 29 arcminute radius around the center of the CTB~102 region at $l = 93\fdg06$, $b = 2\fdg81$. No additional constraints were used in the search and we retrieved 14814 sources. The AllWISE catalog also provides the nearest coincident source ($\leq 3\arcsec$) from the Two Micron All Sky Survey (\emph{2MASS;} \citealt{2mass}) All-Sky data release. This provides data in the J, H, and K\textsubscript{S} bands with magnitude limits at approximately 16.0, 15.0, and 14.5 magnitudes respectively \citep{2mass}. 

\section{Analysis} \label{analysis}

\subsection{The CTB 102 Molecular Clouds} \label{sub:overview}

We adopt the distance to CTB~102 of 4.3~kpc from \cite{arvidsson09}. This is essentially a kinematic-based distance that accounts for known non-circular motions in the second quadrant of the Galaxy \citep{bb93}. Assuming this, we estimate the overall size of the molecular cloud associated with the region to be approximately $60 \times 35$ pc. \autoref{fig:12CO} shows the \co emission integrated over $-71$ km s$^{-1} < V_{\mathrm{LSR}} < -53$ km s$^{-1}$, and indicates the extent of four distinct subdivisions of the molecular cloud, which were identified from visual inspection of the data cube. The semi-major and semi-minor axes of region 1 measure approximately $14 \times 8$ pc, 2a: $9 \times 6$ pc, 2b: $10 \times 9$ pc, and 2c: $11 \times 8$ pc (for more precise numbers see \autoref{tab:properties}). We also distinguish the region 1a ($5 \times 5$ pc), contained entirely within region 1, due to the high concentration of YSO candidates located within this portion of region 1 (see \autoref{fig:RGB}). Finally, we define a region 2 ($28 \times 16$ pc) encompassing regions 2a, 2b, and 2c.

The mass of the molecular clouds associated with CTB~102 can be measured by using the X\textsubscript{CO} factor to convert the integrated \co intensity into an H$_2$ column density, then integrating the column density over the cloud area. Column density was calculated using
\begin{equation}
\label{eqn:XFactor12}
N(\mathrm{H}_{2}) = 2.3 \times 10^{20} \int T^{\mathrm{12CO}}_{\mathrm{MB}}dV \mathrm{cm}^{-2} ,
\end{equation}
where we have used the Milky Way average X\textsubscript{CO} = 2.3 $\times 10^{20}$ cm$^{-2}$ (K km s$^{-1}$)$^{-1}$, which is a robust value applicable to clouds found in a wide range of Galactic environments \citep{bolatto13,szucs,gong18}.  In addition to a statistical uncertainty of $\pm 0.3$ dex \citep{bolatto13}, X\textsubscript{CO} is known to vary systematically with Galactic environment due to changes in both metallicity and local star formation rate, and it is important to consider how significantly the environment around CTB~102 may differ from the Galactic average.

X\textsubscript{CO} is known to increase with decreasing metallicity, and a very sharp upturn is observed for metallicities below 0.5~Z$_\sun$ (see e.g. Figure 9 in \citealt{bolatto13}). For R$_\sun = 8.5$~kpc (8 kpc), CTB 102 has a Galactocentric distance ($R_G$) of 9.7~kpc (9.3~kpc), so $\Delta R_G = R_G-R_\sun \approx 1.2$ kpc (1.3 kpc). The metallicity gradient in the Milky Way is approximately $-0.05$ dex/kpc (Balser et al. 2012); this results in an essentially negligible decrease in metallicity to 0.9~Z$_\sun$ at the distance of CTB~102. In contrast, a hypothetical far outer Galaxy molecular cloud at $R_G = 16$~kpc would have, using the same metallicity gradient, Z $\approx 0.5$ Z$_\sun$, and the Galactic average X\textsubscript{CO} would clearly not be appropriate.

The radiation environment of portions of the CTB~102 molecular clouds clearly deviate from the Galactic average interstellar radiation field (ISRF) due to the proximity of the \hydro region. To roughly quantify this, Cloudy 17.00 \citep{cloudy17} was used to model the radiation field surrounding an O5~V star. The energy density of the radiation field incident on a molecular cloud located at a distance of 1, 3, and 10 pc is $\sim$ 100, 10, and 1 eV~cm$^{-3}$. So, while an extremely strong ISRF is limited to the very close proximity of the O star, increases in the strength of the ISRF by factors between 1 -- 10 are probably common in the clouds surrounding the \hydro region. The effect on changing the ISRF strength on X\textsubscript{CO} has been investigated via simulations. \citet{szucs} models show that X\textsubscript{CO} is only weakly sensitive to the ISRF strength at Z$\sim$Z$_\sun$ (see their models (d), (e) and (j) corresponding to an ISRF of strength $G_0$ = 1, 10, and 100). \citet{clark15} investigated the effect of increased star formation rate, which was modeled as an increase in both the ISRF strength and the cosmic ray ionization rate. They found that X\textsubscript{CO} does increase with increased SFR, but, for clouds with properties probably closest to ours ($M \sim 10^4$ M$_\sun$, $n_o \sim 100$ cm$^{-3}$) the dependence on SFR is weak (a factor of a “few” increase over two orders of magnitude increase in SFR). \citet{szucs} provide a succinct summary of the relationship between X\textsubscript{CO} and SFR: at best the increase is “sub-linear”, and increased density and velocity dispersion within the molecular clouds can act to offset the increase associated with increasing SFR.

Elliptical annuli were used to define the extent of each cloud and to calculate an average background level. A mean molecular weight of 2.3 was used for the total mass calculation. The systematic uncertainties related to ISRF strength and metallicity, which we think are minimal in the case of CTB~102, will tend to make our mass estimates underestimates. Results are shown in \autoref{tab:properties} along with the results of other mass calculation techniques described in this subsection.

We can also estimate the H$_2$ column densities using the integrated map for \textsuperscript{13}CO. Following the procedure described in \cite{Rohlfs}, the \coiso column density is related to the integrated \coiso intensity by:
\begin{equation}
\label{eqn:Xfactor13}
N(^{13}\mathrm{CO}) = 7.3 \times 10^{14} \int T^{\mathrm{13CO}}_{\mathrm{MB}}dV \mathrm{cm}^{-2} .
\end{equation}
In applying \autoref{eqn:Xfactor13} we are assuming that the \coiso emission is optically thin and that LTE applies. We also assume identical excitation temperatures for \co and \coiso of 10~K \citep{szucs,simonBania}. We adopt the conversion factors \co$/$\coiso $= 6.21 \times D_{\mathrm{GC}}(\mathrm{kpc}) + 18.71$ and \co$/$H\textsubscript{2} $= 8 \times 10^{-5}$ from \cite{MILAM05} and \cite{Blake}, respectively, assuming $D_{\mathrm{GC}} = 10.1$ kpc from \cite{arvidsson09}. 

The column density derived using \autoref{eqn:Xfactor13} is sensitive to the excitation temperature chosen; changing the excitation temperature to 20 or 30 K will cause an increase of 40\% or 92\%, respectively \citep{simonBania}. This systematic uncertainty likely dominates over the error in background determination for the regions, which we estimate to be $\sim$10\%. Given this large systematic uncertainty, along with other omitted potential corrections, such as subthermal excitation of higher rotation levels and moderate opacity effects, we consider the \coiso mass estimate to be a lower limit accurate to within a factor of a few \citep{simonBania, arvidsson09, szucs}.

Finally, the viral mass of a molecular cloud is often a good estimate of the true mass of the cloud even when the gas is not truly in virial equilibrium. \cite{szucs} found that, in the context of their numerical models of molecular clouds, the underlying assumption of this mass determination technique is that the observed line widths are dominated by turbulent broadening rather than thermal broadening. The virial masses were determined using 
\begin{equation}
\label{eqn:Virial}
M_{\mathrm{vir}} = 1040 \times R_{pc} \times \sigma_{v}^2 , 
\end{equation}
where $\sigma_{v}$ is the \co velocity dispersion in km s$^{-1}$, and $R_{pc}$ is the geometric mean size (Equation 12 in  \citealt{szucs}). The values used for each region are given in \autoref{tab:properties}.

\begin{deluxetable}{lccccccc}
\tablecaption{Properties of the CTB~102 molecular clouds \label{tab:properties}} 
\tablewidth{0pt}
\tablehead{  Region & 1 & 1a &  2 &  2a & 2b & 2c & Total     }
 \startdata
Size (pc)\tablenotemark{a} & 14.3 $\times$ 8.8 & 5.2 $\times$ 5.0 & 28.1 $\times$ 13.9 & 8.5 $\times$ 5.7 & 10 $\times$ 8.9 & 11.3 $\times$ 8.0 & 30 $\times$ 17.5 \\
$R_{pc}$\tablenotemark{b}  & 11.3  & 5.1 & 19.8 & 7.1 & 9.4 & 9.5 & \nodata \\    
    \hline
    \co Avg. $V$\textsubscript{LSR} (km s$^{-1}$) & $-$62.56 & $-$63.21 & $-$62.55 & $-$63.29 & $-$62.34 & $-$62.10 & $-$62.61\\
    \hspace{8mm} $\sigma_{v}$ (km s$^{-1}$) & 1.72 & 1.44 & 1.53 & 1.56 & 1.31 & 1.60 & 1.85 \\
   \hspace{8mm} Mass (log M\textsubscript{$\odot$}) & 4.27 & 3.81 & 4.62 & 4.00 & 4.06 & 4.14 & 4.78 \\
    \hline
    \coiso Avg. $V$\textsubscript{LSR} (km s$^{-1}$) & $-$64.46 & $-$64.47 & $-$64.45 & $-$64.44  & $-$64.13 & $-$64.19 & $-$64.57\\
    \hspace{8mm} $\sigma_{v}$ (km s$^{-1}$) & 1.20 & 1.16 & 1.23 & 1.03 & 1.29 & 1.15 & 1.22 \\
   	\hspace{8mm} Mass  (log M\textsubscript{$\odot$}) & 3.28 & 2.64 & 3.99 & 3.52 & 3.56 & 3.61 & 4.07\\  
   	\hline
   	Virial Mass (log M\textsubscript{$\odot$}) & 4.54 & 4.04 & 4.68 & 4.26 & 4.22 & 4.40 & 4.95\tablenotemark{c} \\
\enddata
    \tablenotetext{a}{Semi-major and semi-minor axes of the elliptical apertures used to determine the mass of the region from the integrated \co and \coiso data cubes}
    \tablenotetext{b}{Geometric mean size used in the virial mass calculation}
    \tablenotetext{c}{Mass determined from the sum of regions 1 and 2}
 \end{deluxetable}

\subsection{YSO Population of the CTB 102 Molecular Clouds} \label{contamination}

We used \WISE observations of the region to explore the YSO content of the molecular clouds. There are multiple types of sources that can mimic the colors of YSOs, so we must first take steps to remove these contaminants to determine the actual number of YSO candidates in the region. We use $w1, w2, w3$, and $w4$ to refer to the \WISE 3.4 4.6, 12 and 22 $\micron$ band magnitudes from the AllWISE catalog respectively.

To identify likely background star-forming galaxies (SFG) we use the color cuts from \citet[][K17 hereafter]{K17}. These cuts follow \citet[][KL14 hereafter]{Koenig14}, but have more conservative magnitude cuts to reduce the number of spurious transition disk source candidates:
\begin{eqnarray} 
\label{eqn:SFG}
w2 - w3 &>& 1.0  \nonumber \\
w1 -w2 &<& 1.0  \nonumber\\
w1 - w2 &<& 0.46 \times (w2 - w3) - 0.466  \nonumber\\
w1 > 12.0 &\textrm{ or }& w2 > 11.0.
\end{eqnarray}

Active galactic nuclei (AGN) with unresolved broad-line regions are another source of false YSOs as their mid-infrared colors are very similar. These AGN however are expected to be fainter than a typical YSO closer than $\sim$5 kpc. The criteria used to identify and remove these likely AGNs is again identical to K17: 
\begin{eqnarray} 
\label{eqn:AGN}
w1 &>& 1.8 \times (w1 - w3) + 4.1 \nonumber \\
w1 &>& 13.0 \textrm{ or } w2 > 12.0 \nonumber \\
\mathrm{or} \nonumber \\
w1 &>& w1 - w3 + 11.0.
\end{eqnarray}

After the elimination of the red extragalactic contaminants, there are two sources of galactic contaminants that we consider. We once again follow the method from K17 to determine whether any of the objects are unresolved shock emission knots or resolved PAH emission. For shock emission we follow the criteria:
\begin{eqnarray}
\label{eqn:Shock}
w1 - w2 &>& 1.0 \nonumber \\
w2 - w3 &<& 2.0,
\end{eqnarray}
and for PAH emission we use: 
\begin{eqnarray} 
\label{eqn:PAH}
w1 - w2 &<& 1.0 \nonumber \\
w2 - w3 &>& 4.9 \nonumber \\
\mathrm{or} \nonumber \\
w1 - w2 &<& 0.25 \nonumber \\
w2 - w3 &>& 4.75.
\end{eqnarray}

\begin{figure}	
\plotone{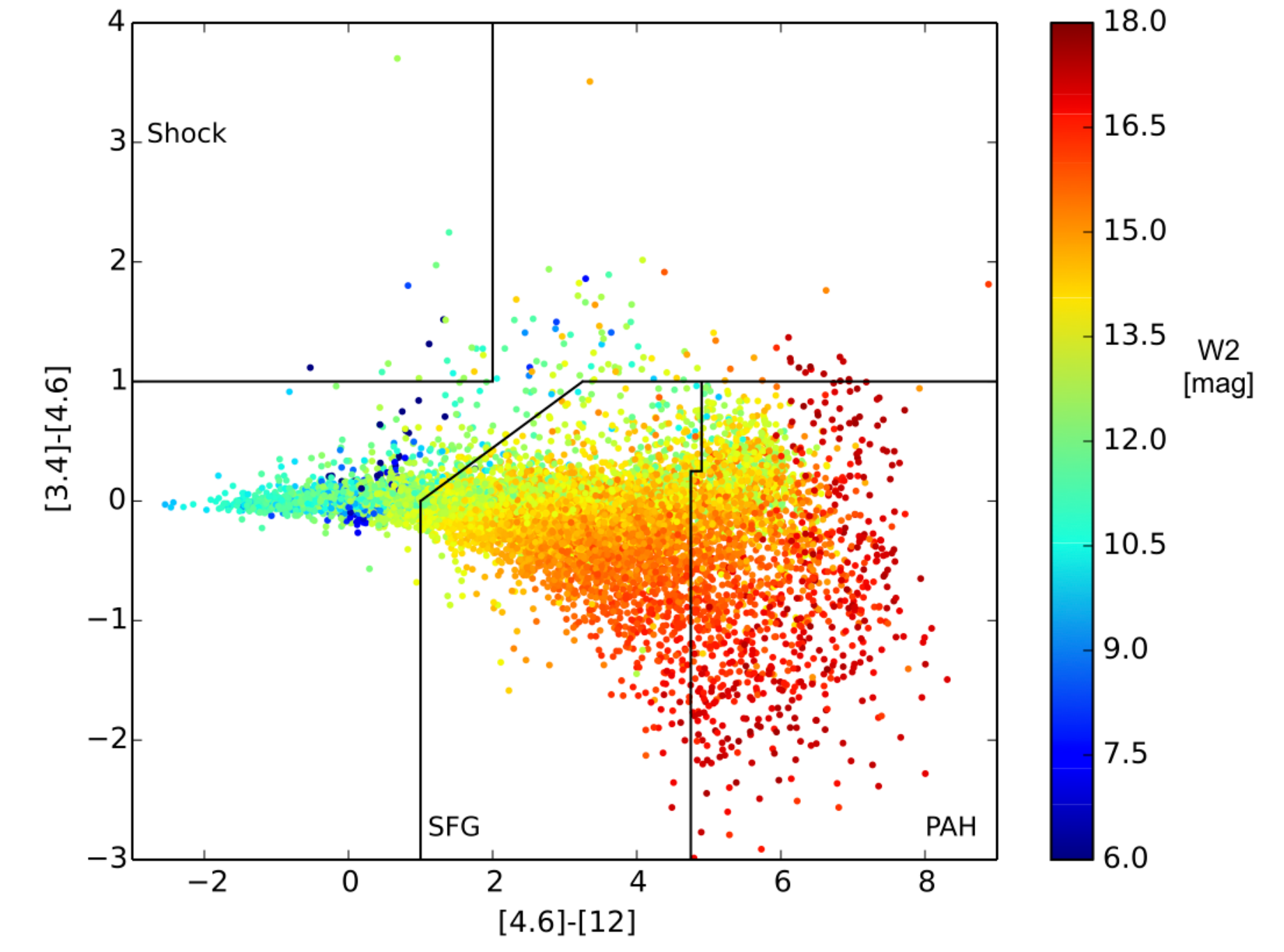}    
\caption{\WISE color-color diagram for \WISE 3.4, 4.6, and 12 $\micron$ bands showing the different contamination cuts described in \autoref{contamination}. All 14814 sources extracted from the AllWISE catalog using a spatial search around CTB~102 are plotted. Sources categorized as shock objects, star forming galaxies and PAH emission are found within the boundaries of the black lines on the plot. Source color refers to the \WISE 4.6 $\micron$ band magnitude. \label{fig:contamination}}
\end{figure}

\autoref{fig:contamination} illustrates the contaminant identification procedure. Starting with the 14814 sources selected from the AllWISE catalog, 13223 sources are identified as contaminants, and the remaining 1591 sources move on to the YSO candidate classification process described in \autoref{classification}.

\subsubsection{YSO Classification} \label{classification}
After removing all of the likely contaminants, we are ideally left with just field stars and YSO candidates. To identify and classify YSO candidates we followed the KL14 procedure. For convenience we show the YSO classification scheme below, and refer the reader to KL14 for details on how the color cuts were developed. Class I objects, essentially YSOs with significant infalling envelopes of material, are defined by: 
\begin{eqnarray}
\label{eqn:ClassI}
w2 - w3 &>& 2.0 \nonumber \\
w1 - w2 &>& -0.42 \times (w2 - w3) + 2.2 \nonumber \\
w1 - w2 &>& -0.46 \times (w2 - w3) - 0.9 \nonumber \\
w2 - w3 &<& 4.5.
\end{eqnarray}

Class II objects, corresponding to YSOs with a significant amount of material in a circumstellar disk, are defined by: 
\begin{eqnarray}
\label{eqn:ClassII}
w1 - w2 &>& 0.25 \nonumber \\
w1 - w2 &<& 0.9 \times (w2 - w3) - 0.25 \nonumber\\
w1 - w2 &>& -1.5 \times (w2 - w3) + 2.1 \nonumber\\
w1 - w2 &>&0.46 \times (w2 - w3) - 0.9 \nonumber\\
w2 - w3 &<& 4.5.
\end{eqnarray}

Transition disk objects, which perhaps are a more evolved Class II object, are defined by:
\begin{eqnarray}
\label{eqn:Transition}
w3 - w4 > 1.5 \nonumber \\
0.15 < w1 - w2 < 0.8 \nonumber \\
w1 - w2 > 0.46 \times (w2 - w3) - 0.9 \nonumber \\
w1 \leq 13.0.
\end{eqnarray}

As most of the sources have corresponding \emph{2MASS} counterparts, KL14 also developed techniques to identify Class I and Class II objects using \emph{2MASS} and \WISE data for sources not classified using only \WISE photometry. \emph{2MASS} Class I objects are defined by:
\begin{equation}
\label{eqn:MassClassI}
H - K\textsubscript{S} > -1.76 \times (w1 - w2) + 2.55.
\end{equation}
\emph{2MASS} Class II YSOs are defined by:
\begin{eqnarray}
\label{eqn:MassClassII}
H - K\textsubscript{S} &>& 0.0 \nonumber \\
H - K\textsubscript{S} &>& -1.76 \times (w1 - w2) + 0.9 \nonumber \\
H - K\textsubscript{S} &<& (0.55/0.16) \times (w1 - w2) - 0.85 \nonumber \\
w1 &\leq& 13.0.
\end{eqnarray}

\begin{figure}	
\plotone{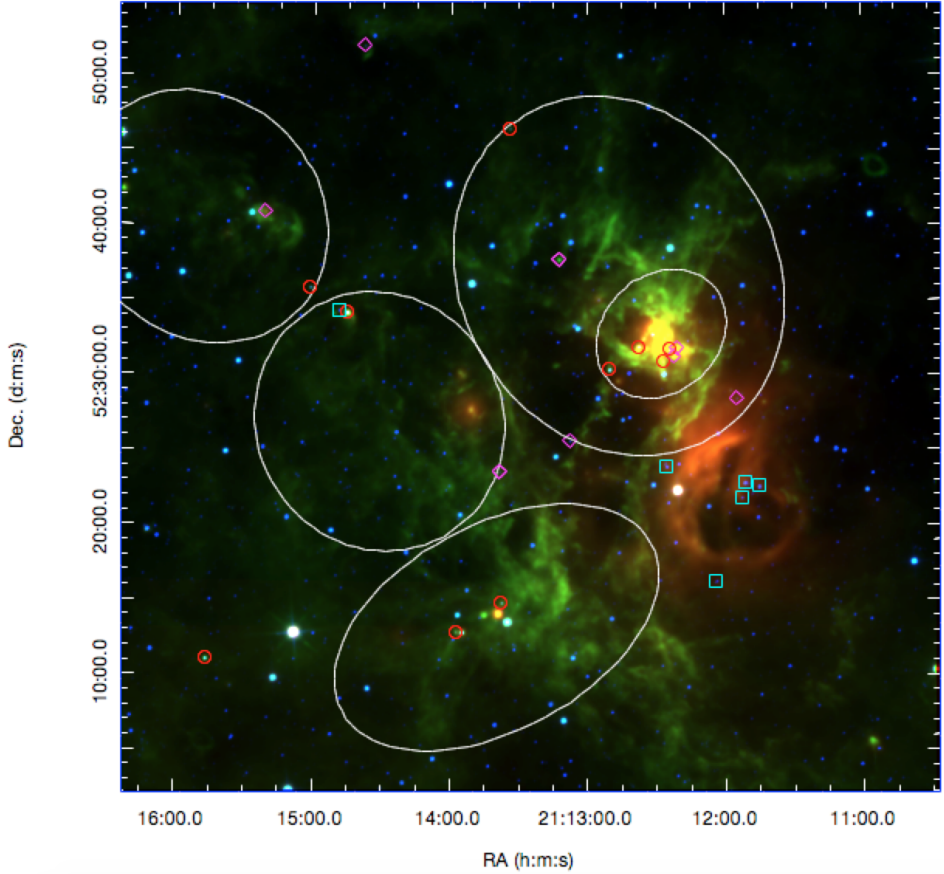}
    \caption{RGB image of CTB102 in \WISE 24, 12, and 4.6 $\mu$m. Red circles correspond to class I sources, magenta diamonds to class II, and cyan boxes to transition disks. Source type comes from the K17 classification given in \autoref{tab:SEDs}. Also shown, as white ellipses, are the molecular cloud regions identified in \autoref{fig:12CO}.}
    \label{fig:RGB}
\end{figure}

We are left with the choice whether or not to consider the $\chi^2$ and signal-to-noise cuts from KL14, which are designed to minimize the number of fake sources that could be potentially classified as YSO candidates. The cuts are the following:
\begin{enumerate}
    \item[]\WISE band 1 (3.4 $\micron$): non-null value for $w1sigmpro$ and $w1chi2$ $< (w1snr - 3)/ 7$,
    \item[] \WISE band 2 (4.6 $\micron$): non-null $w2sigmpro$,
    \item[]\WISE band 3 (12 $\micron$): $w3snr$ $\geq 5$ and either $w3chi2$ $< (w3snr - 8)/8$ or $0.45 < w3chi2 < 1.15$,
    \item[]\WISE band 4 (22 $\micron$): non-null $w4sigmpro$ and $w4chi2$ $< (2 \times w4snr-20)/10$,
\end{enumerate}
where $snr$, $chi2$, and $sigmpro$ are the source signal-to-noise, $\chi^2$, and photometric uncertainty respectively.

Each band cut is applied to the source as that band is required for the classification. For example, only transition disk objects must pass the \WISE band 4 cut, as these are the only object that require $w4$ to be classified as such. 

When following the exact KL14 technique we are left with 18 YSOs. However a visual inspection of the region shows that 11 of those sources do not appear to be reliable, leaving us with 7 YSO candidates using the KL14 method.

\begin{figure}	
\plotone{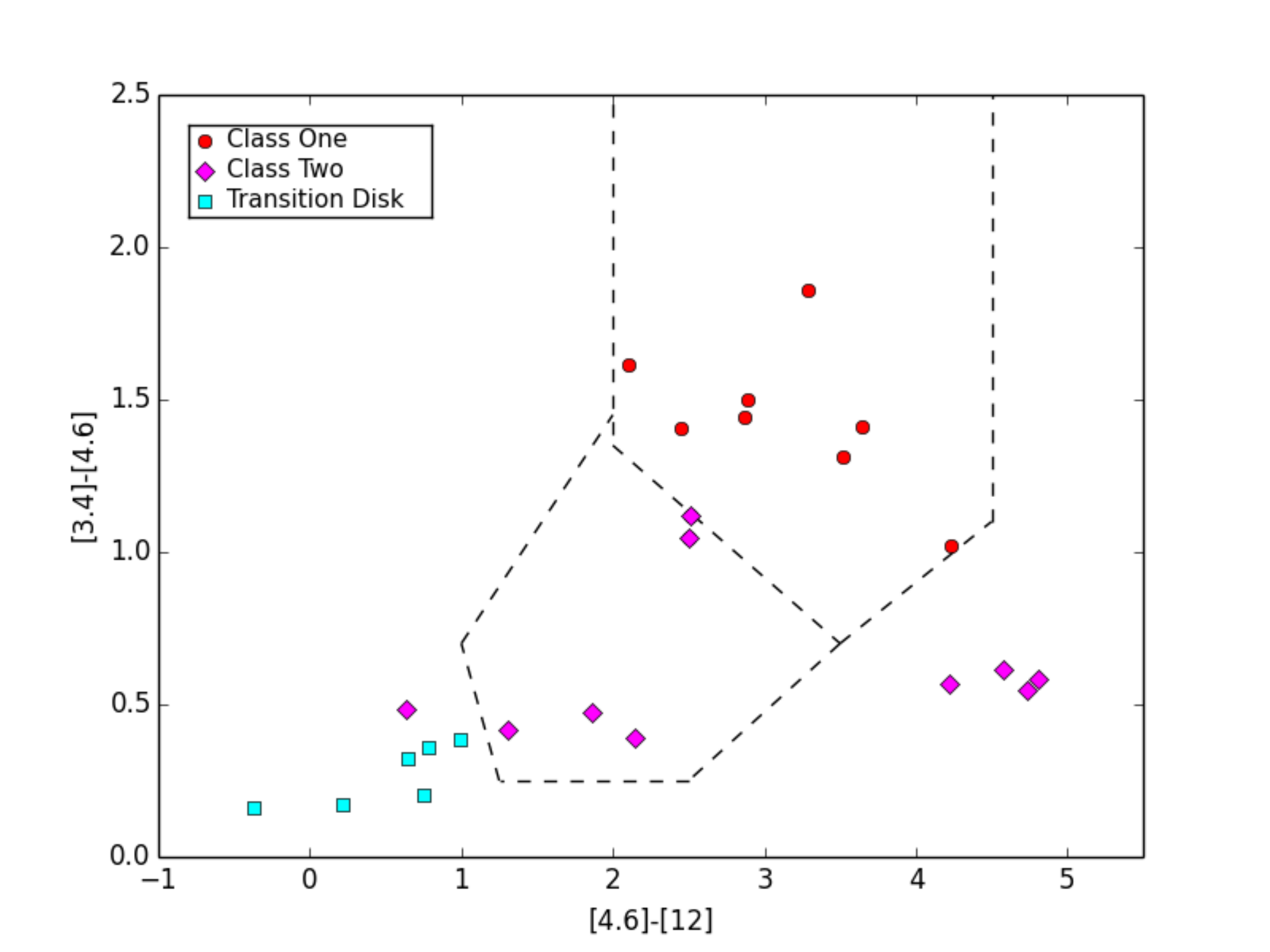}
    \caption{\WISE color-color diagram for \WISE 3.4 4.6 and 12 $\mu$m bands for our \WISE classified YSOs (see \autoref{tab:SEDs}, \autoref{tab:ysophoto}, and \autoref{tab:tdphoto}). The dashed lines denote color criteria for different YSO classes as determined by \WISE photometry. Red sources in the top box are classified as class I YSOs, Magenta sources are class II, and cyan sources are transition disks. Note that the magenta sources not confined to a dashed line region were classified by their \emph{2MASS} photometry. \label{fig:ctbcolorcolor}}
\end{figure}

However if we follow the approach used by K17, we can disregard the $\chi^2$ and signal-to-noise cuts and use a visual inspection to identify spurious YSO candidates after the initial classifications are made. Using this procedure we initially found 128 YSO candidates. The subsequent visual inspection removed 104 sources leaving a total of 24 YSOs (see \autoref{fig:ctbcolorcolor}).

The quality checks from KL14 are very effective at removing spurious sources, and are best used when investigating very large spatial regions. For a smaller area, such as in this study, visual inspection and removal of spurious sources is still practical and results in a better picture of the YSO population. We find 18 Class I and II YSO candidates using this technique, 11 more than from using the KL14 technique. We also find 6 transition disk objects, whereas KL14 identifies none. A number of the transition disk YSOs appear around a 24 $\mu$m bubble-like structure as seen in \autoref{fig:RGB}; however, due to the uncertainty in the evolutionary stage of transition disks, the focus the rest of our study will be on the 18 Class I and Class II YSO candidates (our `best candidate' sample). The YSO classifications for this sample based on \WISE and \emph{2MASS} colors are shown in column 2 of \autoref{tab:SEDs}. Photometry and source designations for this sample and the 6 transition disk candidates are listed in \autoref{tab:ysophoto} and \autoref{tab:tdphoto} in the Appendix. 

\subsubsection{Classification via SED Modeling} \label{sub:SED}

We also classify the YSO candidates in our `best candidate' sample using a slightly modified version of the technique described by \citet{Alexander13}. This technique uses the \cite{Robitaille07} spectral energy distribution (SED) fitting tool to match candidate YSO SEDs to the suite of YSO models from \cite{Robitaille06}. Initially, we attempt to fit reddened stellar models from \cite{Kurucz1993} to the SEDs to ensure that the YSO candidates are not actually reddened stellar photospheres. We allowed A\textsubscript{V} to vary from 0-40 mag.\ and the distance to vary from 4.0 to 4.6 kpc. As in \cite{Alexander13}, we defined a good fit when $\chi^{2}/\mathrm{n_{data}} \leq 2$, where $\mathrm{n_{data}}$ is the number of points in the observed SED. Results of testing each best YSO candidate found that none of the sample SEDs meet this condition, with typical $\chi^{2} \sim 1000$. 

The sources are then matched against the YSO model SEDs using the same fitting tool and the same ranges for interstellar extinction and distance. For each YSO candidate we obtain one or more acceptable fitting models having $\chi^2 - \chi^2$\textsubscript{best}$/n$\textsubscript{data}$< 3$. We then used the criteria from \cite{Robitaille07} to classify each acceptable model based on the values of the mass accretion rate from the YSO envelope ($\dot{M}_{\mathrm{env}}$), the stellar mass ($M_\star$), and the circumstellar disk mass ($M_{\mathrm{disk}}$). Class I YSOs have $\dot{M}_{\mathrm{env}}/M_\star > 10^{-6}$ yr$^{-1}$, Class II YSOs have $\dot{M}_{\mathrm{env}}/M_\star < 10^{-6}$ yr$^{-1}$ and $M_{\mathrm{disk}}/M_{\star} > 10^{-6}$, and Class III YSOs have $\dot{M}_{\mathrm{env}}/M_\star < 10^{-6}$ yr$^{-1}$ and $M_\mathrm{disk}/M_{\star} < 10^{-6}$.

This classification is not always clear-cut since it is possible to find both Class I and Class II type models, or models with different masses, that are acceptable fits to the source SED. For example, in \autoref{tab:SEDs} we see that the best-fit model to source 1983 is a Class II YSO ($\chi^2=2.8$) with a mass of 4.49. There are also 56 other acceptable models (91.8\% of all acceptable models) that are also Class II YSOs, and there are another 5 acceptable models (8.2\% of all acceptable models) that are Class I YSOs. To quantify this inherent fuzziness in the YSO classification we calculate a weighted average class-type and mass as in \cite{Alexander13}. Weights ($P_i$) for each acceptable model ($i$) are calculated using:
\begin{equation}
\label{eqn:weighted}
P_i =  \exp\left(\frac{-\chi^2_i - \chi^2\textsubscript{best}}{2}\right),
\end{equation}
where $\chi^2_i$ is the model $\chi^2$, and $\chi^2_\mathrm{best}$ is the $\chi^2$ for the best-fit model. The weighted average ($\bar{X}$) of parameter $X$ (in our case this is either the model mass or YSO class) is then calculated using: %
\begin{equation}
\label{eqn:average}
\bar{X} = \frac{\Sigma X_iP_i}{\Sigma P_i},
\end{equation} 
where the summation is over all acceptable models.

\begin{figure}[!htbp]
  \centering
  \includegraphics[width=.45\linewidth]{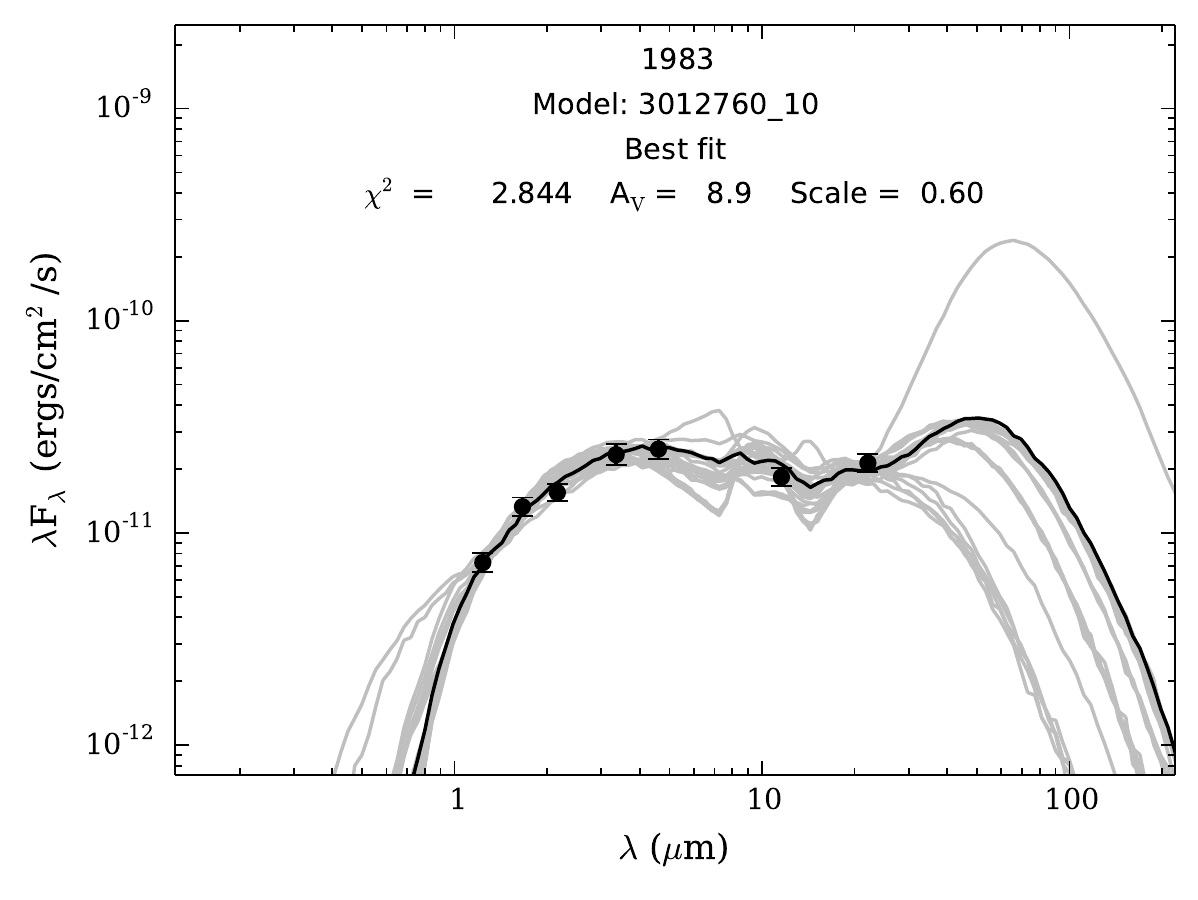}
  \caption{Acceptable model SEDs for source 1983. The best-fit model is shown as a black line. Points are from \emph{2MASS} and \WISE. The best-fit model and \emph{WISE}/\emph{2MASS} photometry both classify the source as a class II YSO. The average class is 1.97 due to the presence of a relatively small number of acceptable class I models.}
  \label{fig:ModelSEDS1983}
\end{figure}

\begin{figure}[!htbp]
  \centering
  \includegraphics[width=.45\linewidth]{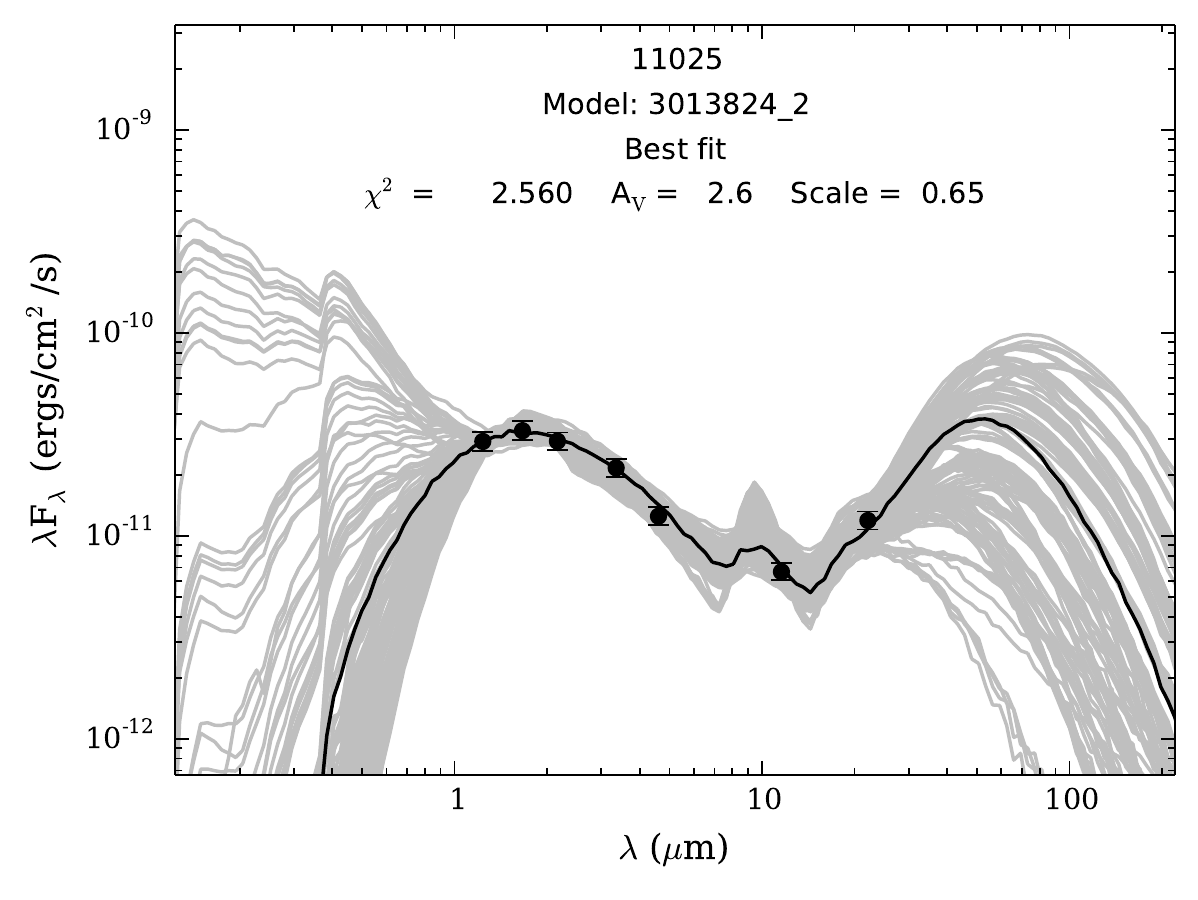}
  \caption{As \autoref{fig:ModelSEDS1983} for source 11025. The best-fit model and \emph{WISE}/\emph{2MASS} photometry both classify the source as a class II YSO. In this case though, the average class is 1.62 as $\sim40$\% of the acceptable SED models are class I.}
  \label{fig:ModelSEDS11025}
\end{figure}

To illustrate the usefulness of procedure we consider two of the YSO candidates, source 1983 and source 11025, in more detail. The best-fit and acceptable model SEDs are shown in \autoref{fig:ModelSEDS1983} and  \autoref{fig:ModelSEDS11025}. In the case of source 1983, we find an average YSO class of 1.97, which reflects the fact that almost all of the acceptable models are Class II. The weighted average mass of 4.41 $M_\odot$ is also similar to the best-fit model mass of 4.49 $M_\odot$. For source 11025, the best SED classification is a class II YSO, but in this case 40\% of the acceptable models have a class I SED. The average YSO class of 1.62 reflects this mix of acceptable models. We also see that the weighted average mass of 4.53 $M_\odot$ is different from the best-fit model mass of 5.45 $M_\odot$. Since the average decimal class (column 8 in \autoref{tab:SEDs}) provides a quantitative measure of the uncertainty of any given classification we prefer to use it over the strict binary class I or class II classification from the \WISE colors.

\begin{deluxetable}{llllcclccccc}
\tablecaption{YSO classifications for best candidate sample \label{tab:SEDs}} 
\tablewidth{0pt}
\tablehead{ & &  \multicolumn{4}{c}{Best Fit} &  \multicolumn{2}{c}{Alternate Fit} &  \multicolumn{3}{c}{Average} & \\
            \cmidrule(lr){3-6}\cmidrule(lr){7-8}\cmidrule(lr){9-11}
\multicolumn{1}{c}{ID}  & \multicolumn{1}{c}{K17}  & \multicolumn{1}{c}{Class\tablenotemark{c}}  & \multicolumn{1}{c}{\% Match\tablenotemark{d}} & Mass\textsuperscript{e} & A\textsubscript{V} & \multicolumn{1}{c}{Class\tablenotemark{f}} & Mass\tablenotemark{g}  &  Class\tablenotemark{h}  & Mass\tablenotemark{i}  & A\textsubscript{V} &Region  \\
    \multicolumn{1}{c}{\#\tablenotemark{a}}  & \multicolumn{1}{c}{Class\tablenotemark{b}} & \multicolumn{1}{c}{($\chi^2$)} & \multicolumn{1}{c}{(\# of Models)} & (M\textsubscript{$\sun$}) & (mag.) & \multicolumn{1}{c}{($\chi^2$)} & (M\textsubscript{$\sun$})  &  & (M\textsubscript{$\sun$})  & (mag.)
    }
 \startdata
	871		&	Class II (W)	&   I (1.5)	&	93.6\% (137) &	4.94 &	4.18 	&	II  (18.7)	&	4.38 	& 1.00 	 & 4.03 & 3.8 & 2b \\	
	1205*	&	Class I	(W)		&	I (5.7)	&	60.0\% (51)	 &	5.15 & 16.6	&	II (15.1)		&	11.7	& 1.01 	&	6.56   & 18.2 &	2b \\
	1983*	&	Class II (W)	&	II (2.8)	&	91.8\% (56)	 &	4.49  &	8.9	& I (8.4)		&	3.60	& 1.97	&	4.41  & 8.4 & 2a \\
	1988	&	Class II (M)	&	II (26)	&	100\% (15)	 &	14.6  & 6.6	&	\nodata			    &	\nodata & 2.00 	&	14.60 & 6.6 &	1  \\
	2927*	&	Class I	(W)		&	II (1.1)	&	91.0\% (496) &	5.14 & 26.3	&I (1.2)		&	6.21	& 1.89	&	5.32   & 27.0 &	2c \\
	3232	&	Class I	(W)		&	I (3.0)	&	69.2\% (72)	 &	1.71 & 15.8 	&	II (4.4)		&	7.45	& 1.43	&	3.84  & 15.7 & 2c \\
	3263*	&	Class II (W)	&	II (2.1)	&	98.2\% (214) &	5.49 & 9.6	&I (16.5)		&	6.56    & 2.00	&	5.74   & 9.3 &	1a \\
	3703	&	Class II (M)	&	II (40)	&	100\% (10)	 &	6.89 & 8.2	&	\nodata				&	\nodata & 2.00 	&	6.89   & 8.2 &	1  \\
	4446	&	Class I	(W)		&	II (7.7)	&	91.9\% (204) &	5.91  &	21.0	&I (15.2)		&	6.70	& 2.00 	&	6.49  & 25.5 & 1a \\
	4692	&	Class II (M)	&	I (13)	&	100\% (4)	 &	7.15  & 1.4	&	\nodata				&	\nodata	& 1.00 	&	7.01  & 1.0 &	2a \\
	5183	&	Class I (M)		&	II (21.8)	&	59.5\% (25)	 &	6.64  &	 1.5 	&I (31)		&	1.10  	& 1.97 	&	6.44 & 2.2 & 1a \\
	5549	&	Class I	(W)		&	II (0.5)	&	91.8\% (529) &	5.49 &	32.3	& I (3.6)		&	0.17	& 1.99	&	5.81   & 31.7 & 1a \\
	5665	&	Class II (M)	&	II (28)	&	20.0\% (11)	 &	6.62 &	11.7	&	I (30)		&	2.89	& 1.77	&	5.82   & 9.3 & 1a \\
	5938	&	Class II (M)	&	I (42)	&	83.1\% (64)	 &	5.62 & 4.1 	&	II (53.3)		&	5.98 	& 1.00  &	5.05  & 3.7 & 1a \\
	6879*	&	Class I (W)		&	II (3.3)	&	95.4\% (267) &	5.26 & 13.8	&I (8)			&	4.92	& 1.99	&	6.24   & 12.1 & \nodata  \\
	7392*	&	Class II (W)	&	II (20.6)	&	82.5\% (14)  &	4.16 & 2.3	&I (35)		&	3.07 	& 2.00	&	4.12   & 2.3 & 1  \\
    8309*	&	Class II (W)	&	I (33)	&	100\% (10)	 &	5.18 & 3.4	&	\nodata				&	\nodata	& 1.00	&	5.19   & 3.4 & 1  \\
   11025 	&	Class II (W)	&	II (2.5)	&	60.0\% (152) &	5.45 & 2.6	&I (4.2)		&	3.14	& 1.62	&	4.53   & 2.9 & \nodata \\
\enddata
    \tablenotetext{a}{All sources were verified by eye; * sources also meet the photometry quality cuts from KL14.}
    \tablenotetext{b}{Source classification based on the criteria in K12, KL14, and K17. W denotes the classification was based solely on \WISE data, and M denotes that \emph{2MASS} and \WISE data were used.}
    \tablenotetext{c}{Source classification based on YSO models best fit to the source SEDs and classification scheme from \citet{Robitaille06} and the $\chi$\textsuperscript{2} of the fit.}
    \tablenotetext{d}{Percentage of good fit YSO models that are the same classification type as the best fit and the number of models that fit.}
    \tablenotetext{e}{Stellar mass of best fit YSO model.}
    \tablenotetext{f}{The classification of the best fit model with a different YSO class than the best fit along with the $\chi^2$ of the fit. In some cases alternate classifications are not present because each well fit model are the same class as the best fit.}
    \tablenotetext{g}{Stellar mass of the alternative classification best fit YSO model.}
    \tablenotetext{h}{Weighted average decimal classification based on the $\chi^2$ for each good fit model.}
    \tablenotetext{i}{Weighted average mass for each good fit model.}
 \end{deluxetable}
 
We see in \autoref{tab:SEDs} that there is a slight trend in the average YSO classification depending on the region where the YSO is located. Regions 1, 1a, 2a, 2b, and 2c have average YSO classifications of 1.77, 1.78, 1.48, 1.33, and 1.66, respectively. We notice that the more evolved YSOs (class II) tend to be in region 1 whereas younger YSOs tend to be found in region 2. This may be indicative of separate (sequential) generations of star formation, where regions 2a, 2b, and 2c are younger because they are located farther from the main \hydro region. We hesitate to make any stronger statement on this point though due to the small sample sizes in regions 2a, 2b, and 2c. The SED analysis was also performed with a fixed distance of 4.3 kpc rather than allowing the distance to vary $\pm .3$ kpc during the fitting. We found that fixing the distance had no appreciable difference in the average mass or class of the YSOs, and therefore would not propagate into our analysis of the star formation efficiency.

\subsection{Star Formation Efficiency} \label{sec:sfe}

The YSO candidates in the CTB 102 molecular clouds detected by \WISE are primarily intermediate- and high-mass objects. Best-fit masses range from 1.71 -- 14.6 M$_\odot$, and weighted masses range from 3.84 -- 14.6 M$_\odot$. To explore if our \WISE data are simply not sensitive to lower-mass YSOs, we used the YSO model photometry from \cite{Robitaille06} to calculate the average apparent magnitude in the \emph{Spitzer} 4.5~$\mu$m band ([I2]) for Class I and Class II YSOs in two mass bins (0.5 -- 1 M$_\odot$ and 1 -- 2 M$_\odot$). Since the \emph{Spitzer} 4.5~$\mu$m and the \WISE 4.6~$\mu$m filters are very similar (e.g., \citealt{Jarrett11}) we assume that the magnitudes are identical for the purpose of this calculation. Each bin contained approximately 1000 models sampling a wide range of disk and envelope properties as well as sampling ten different viewing angles. The average values found for each bin were: Class I (1 -- 2 M $_\odot$) 14.0~(2.3); Class I (0.5 -- 1 M $_\odot$) 14.9~(2.3);  Class II (1 --2 M $_\odot$) 14.4~(1.2); and Class II (0.5 -- 1 M$_\odot$) 15.6~(1.3). The uncertainties, given in the parentheses after the average, were calculated by adding in quadrature the standard deviation due to model variations and the average standard deviation caused by the variable viewing angle. \WISE has a 5$\sigma$ sensitivity at 4.6~$\mu$m of $w2$ $\sim 15.6$ mag \citep{Wright}. Comparing this with the average YSO model magnitudes we see that any observations of sub-solar YSOs will be incomplete, and we conclude that the lack of low-mass YSOs in our sample is most likely an observational limit reflecting the sensitivity of the \WISE data.

To determine the total stellar mass, we can extrapolate our observed sample of YSOs down to sub-solar masses assuming a power-law initial mass function (IMF). Fitting a power law to our sample of 18 YSOs using the weighted average mass values, we obtain $\Gamma{} \equiv d(\log N))/d(\log M) = -1.13\pm 0.99$. Using the best-fit masses, we obtain $\Gamma{} = 0.012\pm 0.32$. Both of these slopes are flatter than the canonical \citet{Salpeter} IMF ($\Gamma{}$ = $-1.35$), but we believe this is due to the small number of YSOs in our sample.

\begin{figure}
    \centering
    \includegraphics[width=.75\columnwidth]{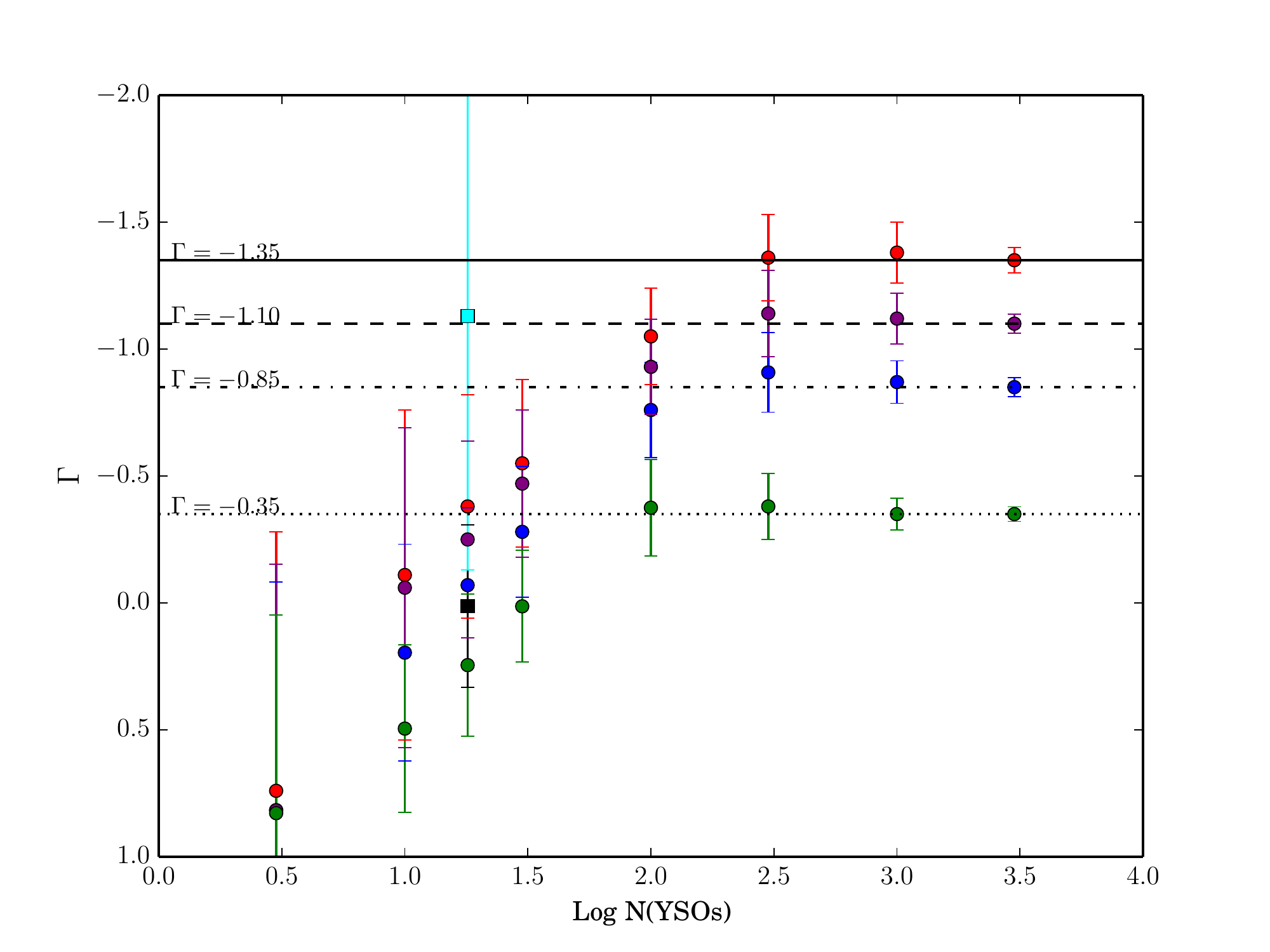}
    \caption{Derived IMF slope $\Gamma \equiv d(\log N))/d(\log M)$ as a function of simulated cluster size for clusters following different underlying IMFs. Red dots correspond to a cluster with an intrinsic $\Gamma = -1.35$, purple follow $\Gamma = -1.10$, blue follow $\Gamma = -0.85$, and green follow $\Gamma = -0.35$. Error bars correspond to the standard deviation of the simulated clusters. The black square corresponds to our observed IMF slope based on the best-fit YSO masses, and the cyan corresponds to the average mass values.}
    \label{fig:monte}
\end{figure}

To explore this further, we designed a simulation to observe how cluster size influences an observed IMF. This was done by simulating clusters with four different underlying IMF slopes, $\Gamma$ = $-$0.35, $-$0.85, $-$1.10, and $-$1.35, and stellar masses ranging between 2 -- 14 M$_\odot$ (the range of mass values in \autoref{tab:SEDs}). For each different cluster IMF, we simulated sample sizes of 3, 10, 18, 30, 100, 300, 1000, and 3000 YSOs. Each simulation was run 1000 times per sample size. We found that the observed IMF slope does not approach the actual value until the sample reaches approximately 100 in size (see \autoref{fig:monte}). For very small YSO sample sizes, it is essentially impossible to get the true IMF slope from the observations. For a Salpeter IMF, we find that a random selection of 18 YSOs will produce $\Gamma{} = -0.38$ $\pm$  $0.44$, matching within the errors of our observed IMF for the best fit YSOs (black square in \autoref{fig:monte}). We also see that our observed IMF is consistent with other power laws within the large standard deviation for small sample sizes, but we find no reason to not follow $\Gamma{} = -1.35$ in estimating the cluster mass in our case. Using $\Gamma{} = -1.35$, we use the 6 -- 7 M$_\odot$ YSOs from \autoref{tab:SEDs} in each region to normalize our IMF, and extrapolate down to the sub-solar mass regime to determine the cluster mass.

The star formation efficiency (SFE), $M_{\mathrm{Cluster}}/(M_{\mathrm{Cluster}} + M_{\mathrm{Gas}})$, was found by considering three different cluster masses. We first look at only our sample of detected YSOs to determine the cluster masses within each region, thus setting a lower limit of the SFE (SFE\textsubscript{Min} in \autoref{tab:SFE}). We also use the Salpeter IMF, with $\Gamma{} = -1.35$, for $0.5~\mathrm{M}_\sun \leq \mathrm{M} < \mathrm{M}_{\sun,\mathrm{Max}} = 7~\mathrm{M}_\sun$ and $\Gamma{} = 0$ for $0.08~\mathrm{M}_\sun <  \mathrm{M} < 0.5~\mathrm{M}_\sun$ (SFE in \autoref{tab:SFE}) . Finally, we derived an alternative cluster mass using the same IMF, but ignoring YSOs with $\mathrm{M} < 0.5~\mathrm{M}_\sun$ (SFE\textsubscript{Alt.} in \autoref{tab:SFE}).

\autoref{tab:SFE} shows our calculated star formation efficiencies for each of the subdivided regions and in total. We remind the reader that systematic effects discussed in \autoref{sub:overview}, which we believe are small in this case, may result in the SFE values being overestimated.

\begin{deluxetable}{lccccccc}
\tablecaption{Cluster Mass and Star Formation Efficiency \label{tab:SFE}}
\tablewidth{0pt}
\tablehead{Region & 1 & 1a &  2 &  2a & 2b & 2c & Total}
 \startdata
    \# of YSOs (Class I and II) & 9 & 6 & 6 & 2 & 1 & 2 & 18\\
    YSO Mass\tablenotemark{a} (M\textsubscript{$\sun$}) 	& 75.2 	 & 35.5   & 27.2   & 11.4   & 6.6    & 9.2    & 108.2  \\
    SFE\textsubscript{Min} 									& 0.4\%  & 0.5\%  & 0.07\% & 0.1\%  & 0.06\% & 0.07\% & 0.18\% \\
    \hline
    Cluster Mass (Log M\textsubscript{$\sun$}) 				& 3.64 	 & 3.52   & 3.5   &  3.20  & 3.20   & 3.19   & 3.87    \\
    SFE $(\mathrm{^{12}CO})$\tablenotemark{b} 							& 19.4\% & 33.9\% & 7.1\%  & 13.7\% & 12.1\% & 10.1\%  & 11.1\% \\
    SFE (virial)\tablenotemark{c} 						& 10.5\% & 23.2\% & 5.8\%  & 8.0\%  & 8.7\%  & 5.8\%  & 7.8\%  \\
    \hline
    Alt. Cluster Mass  (Log M\textsubscript{$\sun$}) 		& 3.5    & 3.34   & 3.32   & 3.03   & 3.03   & 2.68   & 3.70   \\ 
    SFE\textsubscript{Alt} $(\mathrm{^{12}CO})$ 							& 13.7\% & 25.3\% & 4.8\%  & 9.7\%  & 8.5\%  & 3.4\%  & 7.8\%  \\
    SFE\textsubscript{Alt} (virial) 						& 7.4\%  & 16.6\% & 3.9\%  & 5.6\%  & 6.1\%  & 1.9\%  & 5.4\%  \\
\enddata
\tablenotetext{a}{Sum of average YSO masses in the region (see \autoref{tab:SEDs}).}
\tablenotetext{b}{Star formation efficiency based on the cluster mass and the \co mass.}
\tablenotetext{c}{Star formation efficiency based on the cluster mass and the virial mass.}
 \end{deluxetable}

\section{Discussion}\label{summary}

The molecular clouds associated with CTB~102 are likely the substantial remnants of an even larger initial GMC that was the birthplace of CTB~102. They are $60 \times 35$ pc in extent, and our \coiso observations provide a lower limit mass estimate of $10^{4.07}$ M$_\sun$, while our \co observations give a mass range of $10^{4.8} - 10^{5.0}$ M$_\sun$.

We used archival \WISE and \emph{2MASS} photometry to identify and classify 18 class I and class II YSOs within the molecular clouds. SED fitting was used to estimate YSO masses and to refine the YSO classifications. Star formation activity is not spread uniformly throughout the clouds, rather our study reveals pockets of intermediate-mass star formation, closely associated with the high \co and \coiso concentrations seen in \autoref{fig:12CO} and \autoref{fig:RGB}. The lack of observed low-mass YSOs (M $<$ 2 M$_\odot$) is very likely a selection effect due to the sensitivity of the \WISE bands and the distance of the molecular cloud. 

The average YSO classification depends on position within the molecular cloud. We found the average YSO class for region 1 and 2 to be 1.74 and 1.47 respectively. This may be indicative of separate generations of star formation potentially stemming from the O star(s) powering CTB~102.  A stronger statement on the existence of an age/class gradient is precluded because of the small YSO sample size: seven YSOs in region 2 and nine YSOs in region 1. 

A number of transition disk objects were identified, and, although we do not study them in detail, we saw that they appear in a region without any significant amount of \co or \coiso emission. Interestingly, the majority of them instead appear associated with a bubble-like structure seen in the mid-infrared \WISE wavelengths and at 1420 MHz outside of regions 1 and 2c (see \autoref{fig:12CO} and \autoref{fig:1420}). Again, while the sample size is small, the spatial location of these objects, which are presumably more evolved YSOs than class II objects, is consistent with the gradient in age/class mentioned in the previous paragraph.

The total SFE\textsubscript{Min} of the CTB 102 molecular cloud is $0.18\%$, which is quite low. However, this is to be expected, as the sensitivity of \WISE keeps us from observing low-mass YSOs that are likely there. When we use the $M_\mathrm{Cluster}$ values, derived using IMF extrapolations from the observed YSOs (see \autoref{sec:sfe}), the total SFE and SFE\textsubscript{Alt.} values increase and range between $\sim5$ -- $10$\%. This range of SFE is only slightly higher than typical values found for giant molecular clouds (GMCs) at this spatial scale; \citet{Evans2009} find the SFE of molecular clouds on the scale of $10$ -- $70$ pc$^2$ to be between 3 -- 6\%, and \cite{EvansLada1991} found the SFE of L1630 to be $3$ -- $4$\% over a $40 \times 60$ pc region. Given the uncertainties involved with our estimates for the total stellar population, we believe our results for the overall SFE of the CTB 102 molecular clouds are consistent with previous studies of GMCs.

We also examined the SFE of the various sub-regions of the molecular cloud, and found that most regions had SFEs in the 2 -- 10\% range. Region 1a was a clear outlier, with a SFE of 17 -- 35\%, depending on the cluster and gas mass determination technique used. These values are strikingly high for a $\sim5 \times 5$ pc region, and they are more typically associated with the SFE of dense cores on the sub-pc scale. For example, \citet{Lada1992} finds that massive cores in GMCs have a SFE between 30 -- 40\%, and  \citet{Tachihara} calculated the SFE for a sample of 179 cores finding an average SFE of $\sim$10\%, with a values as high as 40\%. We suggest that the 1a region could be an example of a massive embedded protocluster.  These objects have, by definition, a stellar density of $\rho_* > 1$ M$_\sun$ pc$^{-3}$, and region 1a has $\rho_* = 10 - 20$~M$_\sun$~pc$^{-3}$ using a third spatial dimension of 10 -- 5 pc. They also have observed SFEs ranging between 8 -- 33\%, with higher values apparently associated with more evolved clusters \citep{ladalada}. Perhaps the best nearby ($d<2$~kpc) analog, considering both SFE and spatial scale, would be the Mon R2 embedded cluster, which has a 25\% SFE over a 1.85 pc spatial scale (see Table 2 in \citealt{ladalada}).

\section{Conclusions}\label{conclusions}

In this paper we introduced the first high resolution \co and \coiso observations of the molecular clouds associated with the enormous CTB 102 \hydro region, which were obtained using the recently commissioned SEQUOIA-TRAO system. The conclusions of our study are summarized below:

1. We find that what remains of the original molecular cloud has separated into at least two main regions, as seen by the strong divide in the \co and \coiso maps (see \autoref{fig:12CO}), and contains a total mass of $10^{4.8} - 10^{5.0}$ M\textsubscript{$\odot$}.

2. The molecular cloud contains ongoing star formation as seen by the presence of 18 intermediate-mass YSOs within the cloud. When comparing the pockets of star formation, we found that there is a difference in the average YSO class between them, suggesting there are at least two separate generations of star formation.

3. The SFE of the molecular cloud as a whole is comparable with efficiencies found for other Galactic GMCs. However, with a SFE between 17 and 37\%, the region 1a SFE is much higher than what would be expected for a region only $5 \times 5$ pc in size, and we think the region is likely the site of an embedded massive protocluster.

\acknowledgments

We are grateful to the staff members of the TRAO who helped to maintain and operate the telescope and to correlate the data. The TRAO is a facility operated by the KASI (Korea Astronomy and Space Science Institute).

This publication makes use of data products from the Wide-field Infrared Survey Explorer, which is a joint project of the University of California, Los Angeles, and the Jet Propulsion Laboratory/California Institute of Technology, and NEOWISE, which is a project of the Jet Propulsion Laboratory/California Institute of Technology. \WISE and NEOWISE are funded by the National Aeronautics and Space Administration.

This publication makes use of data products from the Two Micron All Sky Survey, which is a joint project of the University of Massachusetts and the Infrared Processing and Analysis Center/California Institute of Technology, funded by the National Aeronautics and Space Administration and the National Science Foundation.

Miju Kang was supported by Basuc Science Research Program through
the National Research Foundation of Korea(NRF) funded by the Ministry of Science, ICT \& Future Planning (No. NRF-2015R1C2A1A01052160).

\appendix 

\WISE and \emph{2MASS} photometric data for our best YSO candiate sample and for the transition disk candidates are provided in \autoref{tab:ysophoto} and \autoref{tab:tdphoto} respectively.

\begin{deluxetable}{lccccccccc}[h]
\tablecaption{Best YSO Candidate Sample \label{tab:ysophoto}}
\tablewidth{0pt}
\centering
\tablehead{WISEA & ID \# & W1 &  W2 &  W3 & W4 & J & H & K & Photometric Quality \\
                 &       & Mag. & Mag. & Mag. & Mag. & Mag. & Mag. & Mag. & Flags}
 \startdata
 	J211338.41+522330.8 & 871	&	10.58	&	10.11	&	8.25	&	4.91	&	12.99	&	11.89	&	11.15	& AAAA AAA 		\\
	J211445.14+523407.2 & 1205	&	9.52	&	7.66	&	4.36	&	1.51	&	18.31	&	15.12	&	12.77	& AAAA UUA 		\\
	J211501.34+523548.3 & 1983	&	10.18	&	9.13	&	6.62	&	4.31	&	14.30	&	12.86	&	11.93	& AAAA AAA 		\\
	J211307.59+522534.4 & 1988	&	9.74	&	9.26	&	8.63	&	7.14	&	12.12	&	11.10	&	10.43	& AAAU AAA 		\\
	J211337.38+521446.8 & 2927	&	10.70	&	9.25	&	6.80	&	3.71	&	16.38	&	14.50	&	13.19	& AAAA UUA 		\\
    J211357.37+521250.9 & 3232	&	11.29	&	9.98	&	6.46	&	3.67	&	16.58	&	15.47	&	13.49	& AAAA UCA 		\\
	J211250.25+523018.8 & 3263	&	9.10	&	7.98	&	5.46	&	3.28	&	14.23	&	12.05	&	10.60	& AAAA AAA 		\\
	J211312.33+523739.4 & 3703	&	11.01	&	10.43	&	5.61	&	4.08	&	14.93	&	13.86	&	12.84	& AAAA AAA 		\\
	J211237.51+523150.4 & 4446	&	10.11	&	8.70	&	5.05	&	2.29	&	15.53	&	14.35	&	14.02	& AAAA UUB 		\\
	J211521.16+524050.0 & 4692	& 	11.91	&	11.34	&	7.12	&	2.58	&	13.94	&	13.06	&	12.62	& AAAA AAA 		\\
	J211226.93+523051.5 & 5183	&	10.95	&	9.93	&	5.69	&	3.04	&	16.83	&	15.48	&	14.28	& AAAA UBA 		\\
	J211224.12+523139.3 & 5549	&	10.73	&	9.28	&	6.41	&	3.67	&	17.37	&	15.03	&	13.60	& AAAA UUA 		\\
	J211222.15+523113.5 & 5665	&	11.13	&	10.52	&	5.93	&	3.06	&	16.02	&	14.78	&	13.88	& AAAA AAA 		\\
	J211220.54+523147.3 & 5938	& 	10.84	&	10.30	&	5.57	&	2.48	&	16.07	&	14.65	&	13.95	& AAAA AAA 		\\
	J211546.43+521102.3 & 6879	&	9.77	&	8.28	&	5.39	&	3.23	&	16.36	&	14.05	&	11.99	& AAAA UAA 		\\
	J211333.92+524623.4 & 7392	&	10.814	&   9.20	&	7.102	&	4.768	&	14.778	&	13.983	&	12.418	& AAAA UAU 		\\
    J211154.69+522825.5 & 8309	&	10.31	&	9.89	&	8.59	&	4.27	&	12.93	&	11.86	&	11.14	& AAAA AAA 		\\
    J211437.81+525154.1	& 11025 &	10.262	&	9.87	&	7.723	&	4.942	&	12.794	&	11.865	&	11.240	& AAAA AAA 		\\
 \enddata
 \end{deluxetable}
 
 \vspace{1cm}
 
\begin{deluxetable}{lccccccccc}
\tablecaption{Transition Disk Candidate Sample \label{tab:tdphoto}}
\tablewidth{0pt}
\tablehead{WISEA & ID \# & W1 &  W2 &  W3 & W4 & J & H & K & Photometric Quality \\
                 &       & Mag. & Mag. & Mag. & Mag. & Mag. & Mag. & Mag. & Flags}
 \startdata
 	J211448.55+523417.1 &	1299&	8.493&	8.2  &	7.539	&	5.077	&	13.07	&	10.379	&	9.171	& AAAA AAA	\\
	J211225.28+522351.9 &	5151&	8.331&	8.158&	7.945	&	5.144	&	10.559	&	9.55	&	9.009	& AAAA AAA	\\
	J211152.50+522147.8 &	8876&	9.338&	8.95 &	7.959	&	4.873	&	11.693	&	10.729	&	10.082	& AAAB AAA	\\
    J211151.18+522247.2 &	8885&	7.874&	7.549&	6.906	&	4.813	&	10.225	&	9.326	&	8.736	& AAAB AAA	\\
	J211203.64+521614.0 &	9079&	8.958&	8.795&	9.167	&	6.091	&	10.933	&	10.075	&	9.582	& AAAA AAA	\\
	J211144.96+522232.5 &	9697&	8.044&	7.683&	6.898	&	4.172	&	10.071	&	9.221	&	8.657	& AAAA AAA	\\
 \enddata
 \end{deluxetable}
 
\bibliography{ref}{}
\bibliographystyle{aasjournal}

\end{document}